\documentclass[%
reprint,
superscriptaddress,
amsmath,
amssymb,
aps,
floatfix,
]{revtex4-2}
\usepackage{graphicx}
\usepackage{dcolumn}
\usepackage{bm}
\usepackage{placeins}
\usepackage[T1]{fontenc}
\begin{document}
\preprint{APS/123-QED}
\title{Spin wave study of magnetic perpendicular surface anisotropy\\ in single crystalline MgO$\text{/}$Fe$\text{/}$MgO films}
\author{J. Solano}
\affiliation{
Institut de Physique et Chimie des Mat\'eriaux de Strasbourg, UMR 7504 CNRS, Universit\'e de Strasbourg, 23 rue du Loess, BP 43, 67034 Strasbourg Cedex 2, France}
\author{O. Gladii}
\affiliation{Helmholtz-Zentrum Dresden - Rossendorf, Institute of Ion Beam Physics and Materials Research, 01328 Dresden, Germany}
\author{P. Kuntz}
\affiliation{Entroview, 11 rue des math\'ematiques, 38402 Saint Martin d'H\`{e}res, France}
\author{Y. Henry}
\author{D. Halley}
\author{M. Bailleul}
\email{matthieu.bailleul@ipcms.unistra.fr}
\affiliation{
Institut de Physique et Chimie des Mat\'eriaux de Strasbourg, UMR 7504 CNRS, Universit\'e de Strasbourg, 23 rue du Loess, BP 43, 67034 Strasbourg Cedex 2, France}

\begin{abstract}
Broadband ferromagnetic resonance is measured in single crystalline Fe films of varying thickness sandwiched between MgO layers. An exhaustive magnetic characterization of the films (exchange constant, cubic, uniaxial and surface anisotropies) is enabled by the study of the uniform and the first perpendicular standing spin wave modes as a function of applied magnetic field and film thickness. Additional measurements of non-reciprocal spin wave propagation allow us to separate each of the two interface contributions to the total surface anisotropy. The results are consistent with the model of a quasi-bulk film interior and two magnetically different top and bottom interfaces, a difference ascribed to different oxidation states.
\end{abstract}

\maketitle

\section{\label{sec:level1_intro}Introduction}

In recent years, there has been an increased interest in magnetic films with large perpendicular magnetic anisotropy due to their potential to improve the efficiency and non-volatility of spin transfer torque magnetoresistive random access memories~\cite{Meng.2006,Kishi.2008,Ikeda.2010,Kim.2011,Moinuddin.2020}. In this search for new materials, Fe films interfaced with MgO are of particular interest due to their favourable properties: a small magnetic damping, a high tunneling magnetoresistance and a large perpendicular surface anisotropy (PSA) \cite{Johnson.1995,Johnson.1996, Kishi.2008,Maruyama.2009,Ikeda.2010}.

There is a good general agreement between experimental and theoretical investigations on the nature and order of magnitude of this PSA, with values ranging between 0.86 and 3.15 $\text{mJ/m}^{2}$\cite{Ikeda.2010,Maruyama.2009,Shiota.2009,Lambert.2013,Koo.2013,Okabayashi.2014,Shimabukuro.2010,Nakamura.2010,Yang.2011,Hallal.2013, Odkhuu.2016}. These are up to two times larger than usual PSA found at the interfaces between transition metals and heavy metals \cite{Guo.2006,Johnson.1995,Johnson.1996} despite weak spin orbit coupling\cite{Yang.2011}. Theoretical works have attributed this large PSA to the hybridization between interfacial oxygen and iron atoms \cite{Shimabukuro.2010, Nakamura.2010}.

Experimentally, it is very challenging to access the internal magnetic environment of ultrathin films and separate the contributions of the top and bottom interfaces to the total PSA of a film. So far, experimental characterizations \cite{Ikeda.2010,Yakata.2009,Nistor.2010,Yamanouchi.2011,Maruyama.2009,Shiota.2009,Lambert.2013,Koo.2013,Okabayashi.2014} have relied on magnetometry measurements to estimate the surface magnetic anisotropy. This means that additional hypothesis were needed to extract the individual surface anisotropies, including comparison with a reference interface or assumptions regarding possible bulk magnetoelastic contributions.

In this work, we separate the top and bottom perpendicular surface anisotropies of single crystalline MgO$\text{/}$Fe$\text{/}$MgO films resorting exclusively to spectroscopic measurements of inhomogeous magnetization dynamics. To achieve this, we perform a broadband ferromagnetic resonance characterization of a thickness series of films MgO$\text{/}$Fe$\text{(}t\text{)/}$MgO ($t\!=\!10\text{-}30$nm) and combine it with a careful study of non-reciprocal spin-wave propagation. Our results show that the films of the entire series behave as a quasi-bulk film interior with two Fe$\text{/}$MgO interfaces that are not magnetically equivalent. We deduce then different top and bottom surfaces anisotropies that are in good agreement with theoretical calculations for ultra-thin films\cite{Shimabukuro.2010,Nakamura.2010,Yang.2011,Hallal.2013,Odkhuu.2016}. This work not only presents a new characterization methodology, but also provides evidence that the large PSA of the technologically relevant ultra-thin films also exists in the thicker films traditionally used in material science.

\section{\label{sec:level2_FMR}Broadband ferromagnetic resonance}

\subsection{Film growth}

The studied films were grown by molecular beam epitaxy on commercial MgO(001) substrates and consist of the following stacks: substrate/MgO(20)/Fe($t$)/MgO(8)/Ti(4.5) (thicknesses in nm). The MgO buffer film was deposited on top of a polished MgO surface at 550 $\text{\textdegree}$C. The Fe film, with thickness $t$ = 10, 15, 20, 25, 30 nm, was subsequently grown at 100 $\text{\textdegree}$C (stair step structure obtained with a movable shutter). Finally, the sample was annealed at 480 $\text{\textdegree}$C and capped with the MgO and Ti layers, both grown at room temperature. The epitaxial relationship between Fe and MgO is such that the [010] and [100] in-plane directions (magnetic easy axes of the Fe film) are rotated by 45$\text{\textdegree}$ with respect to those of the MgO films (aligned with the edges of the substrate).

The crystalline quality of the samples was confirmed in-situ by RHEED. After growth, an X-ray diffraction study revealed a slight tetragonal distortion of the Fe lattice with respect to the bulk, more precisely a 0.5$\%$ out-plane compression accompanied by a 0.7$\%$ in-plane expansion \cite{Magnifouet.2020}.

\subsection{Vector Network Analyzer-Ferromagnetic Resonance}

\begin{figure}[ht]
\includegraphics[width=86mm]{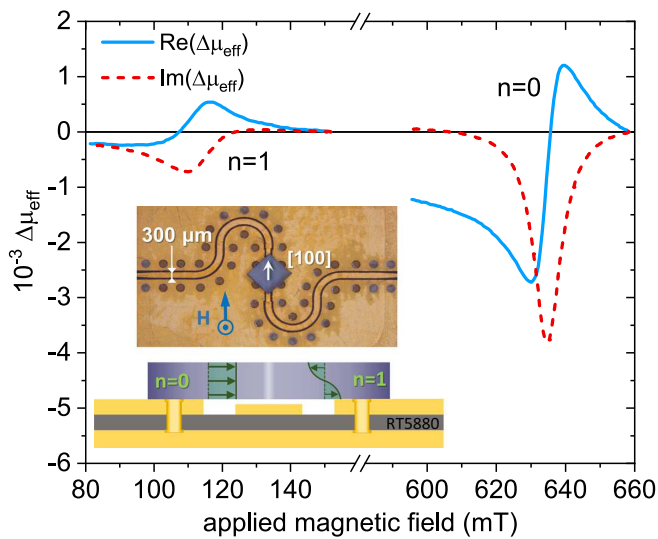}%
\caption{\label{fig:first} Ferromagnetic resonance spectrum measured at 40 GHz for the t=20 nm film. The main panel shows the real and imaginary components of the change of effective permeability as function of the in-plane applied magnetic field (along an the [100] direction of Fe). One distinguishes both a main peak (uniform resonance, n=0) and a satellite one (perpendicular standing spin wave, n=1). The inset contains a photograph of the CPW loaded with a sample at its center (top) and a cross-section sketch showing the tapered CPW, the magnetic film, and the thickness profiles of the dynamic magnetization for the two resonance modes (bottom, not to scale).}
\end{figure}

The dynamic magnetic properties of the films are characterized by Vector Network Analyzer - Ferromagnetic Resonance. The sample ($1.8\!\times\!1.8\text{mm}^2$ piece cut from a film) lies on a 50 $\Omega$ channelized coplanar waveguide (CPW) \cite{R.N.Simons.1989} with a 300 $\mu$m center line separated from the lateral ground planes by 100 $\mu$m gaps (see inset in Fig.~\ref{fig:first}). The 50 $\mu$m thick copper/gold top metallization rests on a 127 µm thick PTFE$\text{/}$glass Rogers RT5880 substrate backed with a very thick copper layer. The CPW's top and bottom grounds are connected through rows of vias parallel to the center line in order to ensure a single-mode propagation in the entire 0-50 GHz frequency range. The part of the waveguide on which the sample is placed has a tapered center line (width 200 $\mu$m, thickness 30 $\mu$m) that compensates for the impedance change caused by the presence of the conductive film on top of the CPW \cite{Bailleul.2013}.

To perform the magnetic measurements, the CPW and sample are inserted in the gap of an electromagnet and connected to a 2-port vector network analyzer via 2.4 mm connectors and coaxial microwave cables. The analyzer can excite and measure the microwave response of the CPW: microwave reflection on each port and transmission between the two ports. The excitation of the ferromagnetic sample results in a modification of the waveguide's impedance. After a suitable calibration and deembedding procedure \cite{bilzer:tel-00202827}, we extract the magnetic field-induced change of the effective permeability $\Delta\mu_{\text{eff}}$ of the waveguide loaded with the ferromagnetic sample. This magnetic response exhibits clear resonances when the microwave frequency matches the field-dependent magnetization precession frequency.

Figure~\ref{fig:first} shows the ferromagnetic resonance spectrum measured on a 20~nm thick Fe film at a microwave frequency of 40~GHz with an external field $\mathbf{H}$ applied in-plane, along the [100] magnetic easy axis (see Fig.~\ref{fig:first} inset). One recognizes an intense peak centered at 634~mT, which we attribute to the uniform resonance mode ($n$=0 in the inset of Fig.\ref{fig:first}). A satellite peak centered at 112 mT is also observed: we attribute this to the first perpendicular standing spin wave mode, corresponding to an inhomogeneous precession across the film thickness, with opposite phases at the two surfaces and zero amplitude at the center ($n$=1 in the inset of Fig.~\ref{fig:first}). The observation of this satellite peak might be surprising at first glance since the microwave magnetic field produced by the coplanar waveguide is expected to be homogeneous over the thickness of the magnetic film, thus preventing the excitation of inhomogeneous modes. However, the fact that the ferromagnetic film is conductive leads to the occurrence of electromagnetic shielding. This effect is characterized by the creation of electrical currents in the metallic film which tend to confine the electromagnetic field within the space between the waveguide and the sample \cite{Bailleul.2013}. This results in a very inhomogeneous microwave magnetic field distribution across the ferromagnetic film thickness leading to the excitation of non-uniform magnetization precession modes \cite{Kennewell.2010}.

\begin{figure*}
\includegraphics[width=120mm]{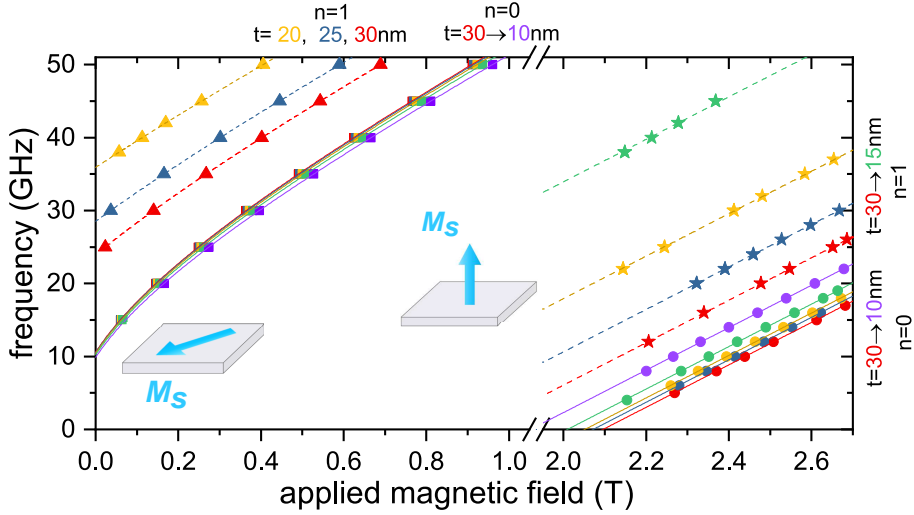}%
\caption{\label{fig:FMRfvsH}
Measured frequency vs. resonance magnetic field of both the uniform mode (n=0) and first perpendicular standing spin wave mode (n=1) for the entire thickness series. In the case of in-plane applied field, squares (triangles) correspond to mode n=0 (n=1). For the case of out-of-plane applied field, circles (stars) correspond to mode n=0 (n=1). The lines are the corresponding Kittel fits [see Eqs. (\ref{eq:p1})].}
\end{figure*}

Similarly, ferromagnetic resonance spectra are recorded for the various thicknesses of Fe with frequency in the range 1.4-50 GHz and external field (up to 2.7~T) applied either in-plane, along the [100] direction of Fe, or out-of-plane (along [001]). Each resonance spectrum is fitted with a complex lorentzian function. From these fits we extract the resonance fields of the two modes, in the two field configurations (Fig.~\ref{fig:FMRfvsH}). From the fit of the resonance peak of mode $n$=0 in the out-of-plane configuration, we also extract the linewidth and from its frequency dependence we finally estimate the damping factor $\alpha$=2.6$\times10^{-3}$. It must be noted that mode $n$=1 could not be observed for $t$=10~nm and 15~nm with in-plane applied field and for $t$=10~nm with out-of-plane field because the corresponding resonance frequencies lie beyond the 50 GHz experimental limit.

\subsection{\label{sec:level2_Model}Theoretical model}
To interpret the resonance data of Figure~\ref{fig:FMRfvsH}, we use the so-called Kittel formulas \cite{Kittel.1948}. These simple expressions are known to be exact in the case of a uniform resonance mode in a high symmetry configuration (equilibrium magnetization parallel or perpendicular to the anisotropy axes). This section shows how they can be extended also to the case of inhomogeneous dynamics ($n\!=\!1$) in films with moderate thickness and sizable surface anisotropies. Our starting point is the linearized Landau-Lifshitz equation for plane spin-waves:
\begin{equation}
\label{eq:LL}
i \omega \textbf{m} = \gamma \mu_o( \textbf{H}_{\text{eq}} \times \textbf{m} -\textbf{M}_{\text{eq}} \times \textbf{h}).
\end{equation}

Here, $\omega$ is the angular frequency, $\gamma$ is the gyromagnetic ratio and $\mu_0$ is the permeability of vacuum. $\textbf{M}_{\text{eq}}$ and $\textbf{m}$ are the static and dynamic components of the magnetization, respectively. Similarly, $\textbf{H}_{\text{eq}}$ and $\textbf{h}$ are the static and dynamic parts of the effective magnetic field, respectively. The effective field derives from the total magnetic energy \cite{Hubert.2014}, which in the present case contains five contributions: i) the exchange and ii) demagnetizing contributions present in any ferromagnet, iii) the cubic volume anisotropy known to exist in iron, iv) surface perpendicular anisotropies at the two Fe/MgO interfaces, and v) an additional volume anisotropy with uniaxial symmetry and perpendicular-to-plane axis which, we argue, is created by strain through a magnetoelastic coupling (see discussion section).

When the external magnetic field $\mathbf{H}$ is applied along the easy axes of the Fe films, the static effective field writes
\begin{equation}
\label{eq:StaticField}
H_{\text{eq}}(\xi)=\begin{cases}
          H \quad &\text{if} \,  $\textbf{H}$ \: \parallel [100]_{\text{Fe}} \\
          H+H_{\text{U}}+h_\text{S}(\xi)-M_\text{s} \quad &\text{if} \, $\textbf{H}$ \: \parallel [001]_{\text{Fe}}, \\
     \end{cases}
\end{equation}

\noindent while the dynamic effective field writes
\begin{equation}
\label{eq:Field}
\textbf{h}= \frac{1}{M_\text{s}} \bigg\{ \left[ \frac{2A}{\mu_0 M_\text{s}}\nabla^2 + H_{\text{K}} \right]\textbf{m} + \left[ H_\text{U}+h_\text{S}(\xi)-M_\text{s} \right]m_{\xi}\boldsymbol{\hat{\xi}} \bigg\}.
\end{equation}

Here, $M_\text{s}$ is the saturation magnetization, $A$ is the exchange stiffness constant, $\boldsymbol{\hat{\xi}}$ is a unit vector along the direction perpendicular to the film and $m_{\xi}$ is the dynamic magnetization component along this direction. Additionally, $H_{\text{K}}\!=\!\frac{2K_1}{\mu_0M_\text{s}}$, where $K_1$ is the volume cubic anisotropy constant, and $H_{\text{U}}\!=\!\frac{2K_\text{U}}{\mu_0M_\text{s}}$, where $K_\text{U}$ is the volume uniaxial magnetoelastic anisotropy constant. Finally, the field $h_\text{s}(\xi)=\frac{2}{\mu_0 M_\text{s}}[K^{\text{bot}}_\text{S} \delta(\xi)+K^{\text{top}}_\text{S}\delta(\xi-t)]$ models the perpendicular surface anisotropies with constants $K_\text{S}^{\text{top}}$ and  $K_\text{S}^{\text{bot}}$ at the top and bottom interfaces, respectively \cite{Gladii.2016p}. Note that $m_{\xi}$=0 when a saturating  magnetic field is applied perpendicular to the plane of the film ($\textbf{M}_\text{eq} \parallel \boldsymbol{\hat{\xi}}$), making Eq.~(\ref{eq:Field}) valid for the two experimental configurations considered in Eq.~(\ref{eq:StaticField}).

The system of equations (\ref{eq:LL}-\ref{eq:Field}) is effectively a fourth order differential equation for the dynamic magnetization with mixed-type boundary conditions (so-called surface pinning), which does not have an exact analytical solution \cite{Gurevich.1990}. However, it is possible to obtain approximate solutions in some limiting cases. In particular, if the exchange energy $A/t$ is much larger than the surface anisotropy $K_\text{S}$, we can expand the dynamic magnetization in a Fourier series of cosine thickness modes (unpinned standing spin wave modes) and, in the spirit of the Kalinikos-Slavin theory of dipole-exchange spin-waves \cite{Kalinikos.1986,Kalinikos.1990}, limit ourselves to the first two terms of the series. Then we may write the complex amplitude of the dynamic magnetization $\textbf{m}^{*}(\xi)$ as
\begin{equation}
\label{eq:Dynamicmag}
\begin{split}
\textbf{m}^*(\xi)=& \quad\: [m^0_{x}\Phi_0+m^1_{x}\Phi_1(\xi)] \: \boldsymbol{\hat{x}} \\ 
&  +  [m^0_{y}\Phi_0+m^1_{y}\Phi_1(\xi)] \: \boldsymbol{\hat{y}},
\end{split}
\end{equation}

\noindent with $x$ and $y$ denoting two directions orthogonal to the static magnetization $\textbf{M}_{\text{eq}}$. $\Phi_0\!=\!1/\sqrt{t}$ is the lowest order term in the Fourier series corresponding to a normalized uniform distribution, and $\Phi_1(\xi)\text{=}\sqrt{2}\cos(\frac{\pi}{t}\xi)/\sqrt{t}$ is the second term corresponding to a normalized non-uniform distribution with a thickness-profile which is antisymmetric with respect to the center of the film (see their sketch in the inset of Fig.~\ref{fig:first}). In the basis of the four orthogonal vector modes  $\left[\Phi_0\mathbf{\hat{x}},\;\Phi_1(\xi)\mathbf{\hat{x}},\; \Phi_0\mathbf{\hat{y}},\; \Phi_1(\xi)\mathbf{\hat{y}}\right]$, the complex amplitude of the dynamic magnetization can be conveniently expressed as $\textbf{m}^*$=$(m^0_{x},m^1_{x},m^0_{y},m^1_{y})$.

After substituting Eqs.~(\ref{eq:Field}) and (\ref{eq:Dynamicmag}) in Eq.~(\ref{eq:LL}), one can project the system of equations (\ref{eq:LL}-\ref{eq:Field}) on this new four-mode basis (see Ref.~\onlinecite{Solano.2017} for a detailed treatment of the projection) and obtain a simplified eigenvalue equation of the form $i\omega\mathbf{m}^*=C\mathbf{m}^*$, where $C$ is the 4$\times$4 dynamic matrix \cite{Gladii.2016p}. The eigenvalues of this matrix are the resonance frequencies. By replacing the eigenvectors of matrix $C$ back in Eq.~(\ref{eq:Dynamicmag}) one may recover the actual oscillation modes of the magnetization.

The dynamic matrix in the presence of surface anisotropies is given in Appendix~\ref{sec:App1} for the case of a spin wave with wavevector $\textbf{k}$ parallel to the $[010]_{\text{Fe}}$ axis. It is important to note that this matrix depends explicitly not only on the total surface anisotropy $K_{\text{S}}\!=\!K_{\text{S}}^{\text{bot}}\!+\!K_{\text{S}}^{\text{top}}$ but also on the difference of surface anisotropies at the two interfaces $\Delta K_{\text{S}}\!=\!K_{\text{S}}^{\text{bot}}\!-\!K_{\text{S}}^{\text{top}}$~\cite{Gladii.2016p}.

Let us now concentrate on the case of ferromagnetic resonance ($k\!=\!0$). Up to first order in $\Delta K_{S}$, our approach produces Kittel-like\cite{Kittel.1948} expressions for the resonance frequencies of the first two standing spin wave modes:
\begin{subequations}
\label{eq:p1}
\begin{eqnarray}
f_{\parallel n}=\frac{\gamma\mu_0}{2\pi}
[(H+H_{Xn})(H+H_{Yn})]^{1/2},\label{subeq:1}
\end{eqnarray}
\begin{equation}
f_{\perp n}=\frac{\gamma\mu_0}{2\pi}
(H-H_{Zn}).\label{subeq:2}
\end{equation}
\end{subequations}

Here $n\!=\!0,1$ is the mode index, and $\parallel$ and $\perp$ refer to the configurations with $\mathbf{H} \parallel [100]_{\text{Fe}}$ and $\mathbf{H} \parallel [001]_{\text{Fe}}$, respectively. $H_{\alpha n}$ are orientation and mode-dependent stiffness fields whose expressions are given in Table~\ref{tab:table1},
\begin{table}[ht]
\caption{\label{tab:table1}%
Explicit expressions of stiffness fields in the resonance frequencies of Eqs.(\ref{eq:p1}).
}
\begin{ruledtabular}
\begin{tabular}{lcc}
\textrm{Field} & \textrm{$n=0$}& \textrm{$n=1$}\\
\colrule
$H_{Xn}$ & $H_{\text{K}}$ & $H_{\text{K}}+H_{\text{E}}$\\
$H_{Yn}$ & $M_{\text{s}}+H_{\text{K}}-H_{\text{U}}-H_{\text{S}}$ & $M_{\text{s}}+H_{\text{K}}-H_{\text{U}}-2H_{\text{S}}+H_{\text{E}}$\\
$H_{Zn}$ & $M_{\text{s}}-H_{\text{K}}-H_{\text{U}}-H_{\text{S}}$ & $M_{\text{s}}-H_{\text{K}}-H_{\text{U}}-2H_{\text{S}}-H_{\text{E}}$\\
\end{tabular}
\end{ruledtabular}
\end{table}
where the following thickness-dependent exchange and surface anisotropy fields have been used:
\begin{subequations}
\label{eq:HeHs}
\begin{eqnarray}
H_{\text{E}}=\frac{2\pi^2A}{\mu_0M_\text{s}t^2},\label{subeq:5}
\end{eqnarray}
\begin{equation}
H_{\text{S}}=\frac{2 K_\text{S}}{\mu_0M_\text{s}t}.\label{subeq:6}
\end{equation}
\end{subequations}

At the chosen level of approximation, which is valid only for small differences in surface anisotropies ($\Delta K_{\text{S}}\: t/A \ll 1$), the resonance frequencies (Eqs.~\ref{eq:p1}) depend on $K_{S}$ but not on $\Delta K_{S}$, although the $C$ matrix depends explicitly on it. This may be explained as follows. With $k$=0, the mutual demagnetization factor $Q$ vanishes (see Appendix~\ref{sec:App1}) and so does the largest source of hybridization between the uniform and antisymmetric basis modes. Only terms in $\Delta K_{\text{S}}$ remain non zero in the off diagonal blocks of the $C$ matrix (Eq.~\ref{eq:Cmatrix}), meaning that the difference in surface anisotropies becomes the sole source of hybridization. It happens that the corresponding coupling is proportional to $\Delta K_{\text{S}}^2$ and is thus neglected in the above approximation. We note also that the contribution of surface anisotropies to the stiffness fields is doubled in the case of mode $n$=1 as compared to mode $n$=0. This is a direct consequence of mode $n$=0 being uniform at this level of approximation while mode $n$=1 is fully asymmetric with large amplitude at the surfaces, which makes it more sensitive to PSA.

\subsection{Results}\label{sec:results}

\begin{figure}[ht]
\includegraphics[width=86mm]{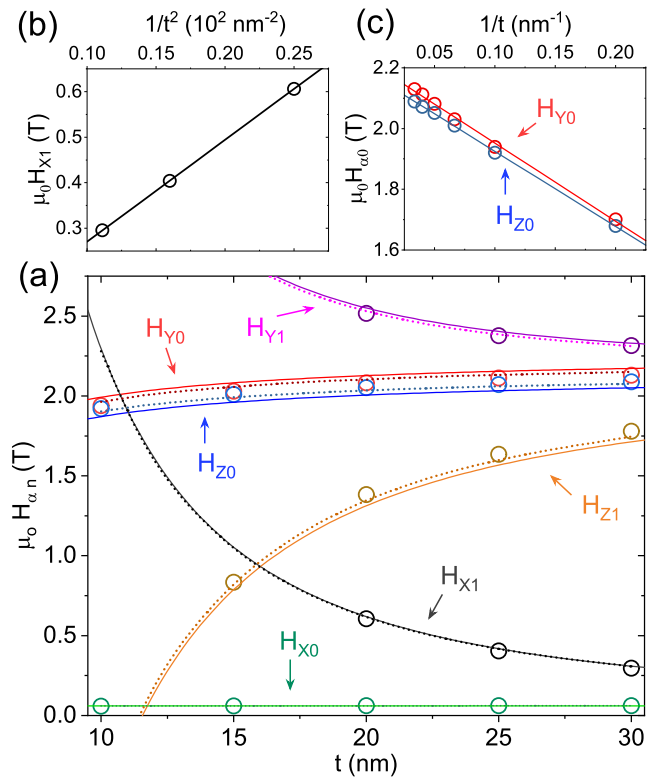}%
\caption{\label{fig:HiFields} a) Stiffness fields as a function of Fe thickness obtained from Kittel-like fits of the experimental resonance frequencies (open circles) compared with the analytical model of Table~\ref{tab:table1} (solid lines) and the values obtained from Kittel-like fits of SWIIM simulations (dotted lines). b) Stiffness field $H_{X1}$ as a function of $1/t^2$. c) Stiffness fields $H_{Y0}$ and $H_{Z0}$ as a function of $1/t$. }
\end{figure}

Fitting the experimental data in Fig.~\ref{fig:FMRfvsH} to the corresponding Eqs.~(\ref{eq:p1}) yields the values of the stiffness fields $H_{Xn}$, $H_{Yn}$, and $H_{Zn}$ $(n\!=\!0,1)$ presented in Fig.~\ref{fig:HiFields} as open circles. For this extraction, we assume a unique value for $\gamma$ (see Table~\ref{tab:table2}),
\begin{table}[b]
\caption{\label{tab:table2}Magnetic parameters obtained for MgO/Fe($t$)/MgO films ($t\!=\!10\!-\!30\text{nm}$).}
\begin{ruledtabular}
\begin{center}
\begin{tabular}{cccc}
$\mu_0 M_{\text{s}}$  [T] & $\gamma/2\pi$ [GHz/T] & $A$ [pJ/m] & $K_1$ [kJ/$\text{m}^3$] \\
\colrule
2.15\footnote{Tabulated value for bulk iron at room temperature} & $29.1\pm0.6$ & $19.4\pm0.1$ & $52\pm1$\\
\hline\\[-1em]
\hline\\[-1em]
$K_\text{U}$ [kJ/$\text{m}^3$] & $K_\text{S}$ [mJ/$\text{m}^2$] &\multicolumn{2}{c}{${\Delta K_{\text{S}}}$ [mJ/$\text{m}^2$] }\\
\colrule
$-45\pm28$ & $2.3\pm0.3$ & \multicolumn{2}{c}{$0.8\footnote{Value obtained from non-reciprocal spin wave measurements (See Section~\ref{sec:level3_PSWS})}\pm0.1$}\\
\end{tabular}
\end{center}
\end{ruledtabular}
\end{table}
which is the average over all film thicknesses of the individual $\gamma$ values obtained by fitting $f_{\perp 0}(H)$ experimental data to Eq.~(\ref{subeq:2}).

Values of the stiffness fields associated with the two oscillation modes $(n\!=\!0,1)$ can be readily treated and combined sequentially so as to extract most of the magnetic parameters of the iron films. As a starting point and in agreement with SQUID characterization, the saturation magnetization value is set to that of bulk iron $\mu_0 M_{\text{s}}\!=\!2.15$~T. Next, we observe that $H_{X0}$ is thickness independent and we extract the cubic anisotropy constant $K_1$ from $H_{X0}\!=\!H_{\text{K}}$. Then, since $H_{X1}\!-\!H_{X0}\!=\!H_{\text{E}}$ varies as $t^{-2}$ [Fig.~\ref{fig:HiFields}(b)], we confidently extract a thickness independent exchange constant $A$ (Eq.~\ref{subeq:5}). We subsequently notice that the thickness dependent parts of $H_{Y0}$ and $H_{Z0}$ both vary as $t^{-1}$, with similar slopes [Fig.~\ref{fig:HiFields}(c)], from which we determine the average total surface anisotropy constant $K_{\text{S}}$ (Eq.~\ref{subeq:6}). Finally, we determine the unixial anisotropy constant $K_{\text{U}}$ using $H_{Y0}\!+\!H_{Z0}\!-\!2(M_{\text{s}}\!-\!H_{\text{S}})\!=\!2H_{\text{U}}$.

The obtained magnetic parameters are summarized in Table~\ref{tab:table2}. In the chosen parametrization, the negative sign of $K_{\text{U}}$ indicates an easy-plane parallel to the film's plane and the positive sign of $K_{\text{S}}$ an easy-axis along the film normal.

The lines in Fig.~\ref{fig:HiFields} are the theoretical stiffness fields calculated by injecting the parameters just determined (Table~\ref{tab:table2}) back into our analytical model. The good agreement obtained illustrates that a single set of thickness independent magnetic parameters is indeed enough to capture most features of the magnetization dynamics in the studied MgO/Fe/MgO films. Furthermore, the fact that the agreement also applies to those stiffness fields which have not been considered in the above analysis ($H_{Y1}$ and $H_{Z1}$) can be considered as a validation of the model. There are however points of slight disagreement between the experimentally determined values and the corresponding analytical predictions. In particular, predicted values of the difference $H_{Y0}\!-\!H_{Z0}$ are significantly larger than the ones determined experimentally. As may be inferred from Appendix~\ref{sec:App3}, modifying the analytical model (Eqs.~\ref{eq:p1}) to include terms up to second order in $\Delta K_{\text{S}}$ allows one partially reducing the disagreement. This second order approximation however produces cumbersome expressions which are unpractical and, even more importantly, unable to provide information regarding the sign of $\Delta K_{\text{S}}$. This points at the need for an accurate determination of this additional parameter through an experimental technique which is sensitive to it at first order, namely spin wave frequency non-reciprocity.

\section{\label{sec:level3_PSWS}Non-reciprocal spin wave propagation}

\begin{figure}[ht]
\includegraphics[width=86mm]{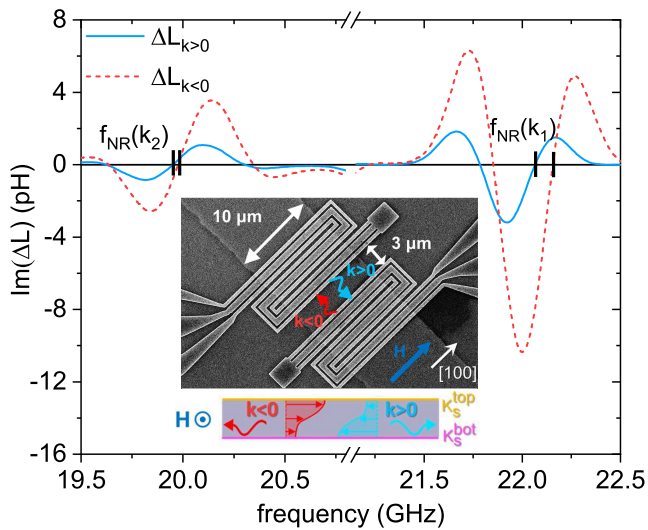}%
\caption{\label{fig:dLs} Measured imaginary component of the mutual inductance between antennas due to counter-propagating spin waves (with wave numbers: $k_1$=3.9 $\text{rad/}\mu \text{m}$ and $k_2$=1.57 $\text{rad/}\mu \text{m}$) in a strip of $t$=20 nm Fe film under a 120 mT magnetic field (Damon-Eshbach geometry). The inset shows an electronic microscope image of the experimental device (top) and a sketch of the asymmetric modal profiles of counter-propagating waves across the thickness of the film and its interplay with asymmetric magnetic surface anisotropies (bottom).}
\end{figure}

Propagating spin wave spectroscopy has been shown to be very sensitive to magnetic asymmetries across the thickness of thin films \cite{Gladii.2016p}, including differences in surface anistropies at the two interfaces ($\Delta K_{\text{S}}\!\neq\!0$). The principle of such measurement is sketched in the inset of Fig.~\ref{fig:dLs}. A spin-wave propagating in the so-called Damon-Eshbach geometry (i.e. with its wave-vector $\mathbf{k}$ oriented perpendicular to the in-plane applied magnetic field  $\mathbf{H}$) is known to exhibit a mode profile non-reciprocity. This means that the wave has an asymmetric distribution across the film thickness, with more amplitude on one side of the film than on the other. This asymmetric profile is reversed when changing the sign of $\mathbf{k}$, i.e. for spin waves propagating in the opposite direction~\cite{Kostylev.2013} (see inset of Fig.~\ref{fig:dLs}). Consequently, an inhomogeneous magnetic environment across the thickness will have different effects on the dynamics of counter-propagating spin waves. This can be measured experimentally as a difference between their resonance frequencies which becomes a spectroscopic signature of the film's asymmetric magnetic environment~\cite{Gladii.2016p}.

To measure this frequency non-reciprocity, suitable devices are fabricated from the Fe samples. This is achieved by patterning the film into a strip geometry and fabricating a pair of conductors on top. These conductors with a meander geometry serve as antennas for exciting and detecting spin-waves of controlled wavelength. With the design chosen in the present work (see inset in Fig.~\ref{fig:dLs}) the most important excitation occurs around two particular wave vectors $k_1$=3.9 $\text{rad/}\mu \text{m}$ and $k_2$=1.57 $\text{rad/}\mu \text{m}$. By using two separate antennas, and interchanging their role, it is possible to measure the changes in mutual inductance corresponding to spin waves propagating with positive and negative wave vectors $\Delta L_{k>0}$, $\Delta L_{k<0}$ \cite{V.Vlaminck.2010}. More details on the fabrication process of the devices and experimental procedure can be found elsewhere \cite{Gladii.2017}.

\begin{figure}[ht]
\includegraphics[width=86mm]{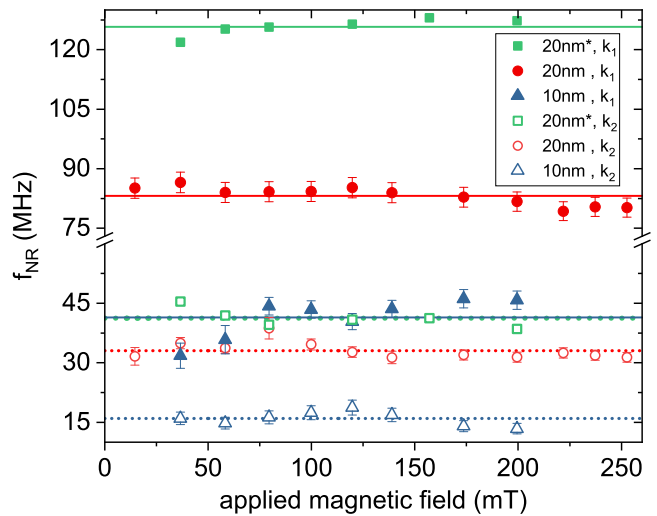}%
\caption{\label{fig:NRvsH} Frequency non-reciprocity measurements as a function of applied magnetic field for devices $t$=10 nm, 20 nm and 20 nm* and each wave numbers $k_1$=3.9 $\text{rad/}\mu \text{m}$, $k_2$=1.57 $\text{rad/}\mu \text{m}$. The lines correspond to the average value of $f_{\text{NR}}$ over the experimental field range.}
\end{figure}

Fig.~\ref{fig:dLs} shows the measured change in mutual inductance corresponding to spin wave propagation in a $t$=20~nm device under a 120~mT field applied along [100]. One can observe directly a frequency difference between counterpropagating waves, both for the main spin-wave excitation [$f_{\text{NR}}(k_1)$] and for the secondary one [$f_{\text{NR}}(k_2)$]. The value of this frequency non-reciprocity is followed as a function of the applied magnetic field in the range 30-200 mT. The different symbols in Fig.~\ref{fig:NRvsH} show the frequency non-reciprocity measured for three different samples: namely devices from the FMR series with $t\!=\!10$~nm and 20~nm, and a third device with $t\!=\!20$~nm but without Ti capping, labelled thereafter 20~nm*. We observe that $f_{\text{NR}}$ is roughly field independent.

To analyze quantitatively the data in Fig.~\ref{fig:NRvsH} and estimate the difference in magnetic surface anisotropy $\Delta K_{\text{S}}$, we use the theory developed in the previous section with some modifications. To account for the propagating character of the spin waves, the space-dependent part of the dynamic component of the magnetization [Eq. (\ref{eq:Dynamicmag})] becomes $\textbf{m}^*(\xi)e^{-i k\eta}$ where $\hat{\eta}$ is a direction vector along the ferromagnetic strip. Now, the dynamic magnetic field writes
\begin{equation}
\label{eq:PropDynH}
\begin{split}
\textbf{h}=& \frac{1}{M_{\text{s}}} \bigg\{ \left[ \frac{2A}{\mu_0 M_{\text{s}}}\nabla^2 + H_{\text{K}} \right]\textbf{m} + \left[ H_{\text{U}}+h_{\text{s}}(\xi) \right]m_{\xi}\boldsymbol{\hat{\xi}} \bigg\}\\
&+ \int_{0}^{t} G_k(\xi-\xi')\textbf{m}\,d\xi',
\end{split}
\end{equation}

\noindent where the last term is the dipolar contribution and $G_k$ the magnetostatic Green's tensor \cite{GUSLIENKO.2011}. As in the previous section, the system of Eqs. (\ref{eq:LL},\ref{eq:PropDynH}) can be projected onto the spin wave modes basis which allows one to obtain an eigenvalue equation with a dynamical matrix $C$ given explicitly in Appendix~\ref{sec:App1} (note that in this case $k\!\neq\!0$). In the Damon-Eshbach configuration, the non-uniform character of magnetization along $\hat{\eta}$ gives rise to dipolar coupling between Fourier components $n\!=\!0$ and $n\!=\!1$ [through the factor $Q\!\neq\!0$ in the non-diagonal blocks of the dynamic matrix $C$, see Eq. (\ref{subeq:3})], which is non-reciprocal with respect to the wave number. This, combined with the asymmetry produced by $\Delta K_{\text{S}}$, is at the very origin of the frequency non-reciprocity.

Once again we can calculate the resonance frequencies from the dynamic matrix C, and in this case, obtain the frequency difference between counterpropagating waves. Assuming $\Delta K_{\text{S}}t/A\!\ll\!1$, an approximate expression can be derived in which the frequency non-reciprocity of mode $n\!=\!0$ is a linear function of both the wave number $k$ and the difference in surface anisotropy $\Delta K_{\text{S}}$ \cite{Gladii.2016p}:
\begin{equation}
\label{eq:LnNonReci}
\begin{split}
f_{\text{NR}} &= f_{\parallel 0}(k<0)-f_{\parallel 0}(k>0) \\
&\approx\frac{8\gamma\:\Delta K_{\text{S}}}{\pi^3M_{\text{s}}(1+\frac{\Lambda^2\pi^2}{t^2})}\:k
\end{split}
\end{equation}

\begin{figure}[ht]
\includegraphics[width=86mm]{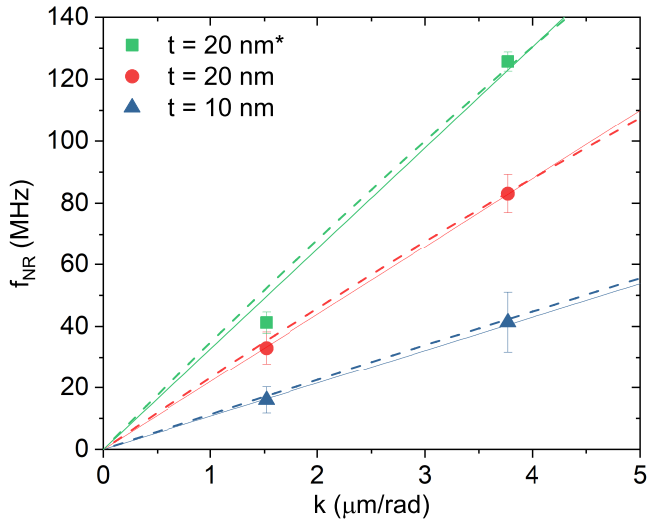}
\caption{\label{fig:NRvsk} Frequency non-reciprocity as a function of wave number for the three films under study. The points represent the average experimental values in the applied field range 30-200 mT (see Figure~\ref{fig:NRvsH}). The solid lines are the corresponding linear fits to Eq. (\ref{eq:LnNonReci}). The dashed lines show the mean value as calculated with SWIIM while using values for $\Delta K_{S}$ that best adjust the experimental points (see text for details).}
\end{figure}

Since $f_{\text{NR}}$ shows no systematic variation with the external magnetic field (Fig.~\ref{fig:NRvsH}), we consider below its average value over the 30-200 mT range and plot it in Fig.~\ref{fig:NRvsk} as a function of $k$ for each of the studied films. Applying Eq. (\ref{eq:LnNonReci}) and using values of $\gamma$, $M_{\text{s}}$, and $A$ found in Table~\ref{tab:table2}, we finally extract estimates for $\Delta K_{\text{S}}$ from the slopes of linear fits: 0.8, 1.1, 1.6 $\text{mJ/m}^{2}$ for the 10 nm, 20 nm and 20nm* films, respectively.

These results confirm the asymmetry of the two film interfaces suggested by FMR measurements. They show that $\Delta K_{\text{S}}$ is undoubtedly positive, which means that, in all films, the bottom interface has a stronger PSA than the top one. Since the above characterization is based on several approximations [Eqs.~(\ref{eq:Dynamicmag}, \ref{eq:LnNonReci})], the magnitude of $\Delta K_{\text{S}}$ should however be refined before comparing and contrasting the films and their respective interfaces. To this aim, we finally turn to numerical simulations.

We resort to the code SWIIM \cite{henry2016propagating} which provides a finite-difference numerical solution of Eqs.~(\ref{eq:LL},\ref{eq:PropDynH}) to calculate the difference between the frequencies of counterpropagating waves as a function of wave number $k$ in the Damon-Eshbach configuration. For each sample, we adjust $\Delta K_{\text{S}}$ (the remaining magnetic parameters are taken from Table~\ref{tab:table2}), so as to best reproduce the experimental $k_{\text{NR}}(k)$ data in Fig.~\ref{fig:NRvsk} (dashed lines). The values of $\Delta K_{\text{S}}$ obtained in this way are 0.7, 0.8, and 1.2 $\text{mJ/m}^{2}$ for the 10~nm, 20~nm and 20~nm* films, respectively. Comparing them with values obtained from Eq.~(\ref{eq:LnNonReci}), we observe that the analytical approach systematically underestimates the effect of $\Delta K_{\text{S}}\!\neq\!0$. Noticeably, using numerical simulations allows us reducing the difference between the values of $\Delta K_{\text{S}}$ for the 10~nm and 20~nm samples to almost nothing, which is of course expected for films of similar composition. We then choose the value $\Delta K_{\text{S}}\!=\!0.8$~mJ/m$^{2}$ as a reference for our MgO/Fe/MgO system.

Having refined the magnitude of $\Delta K_{\text{S}}$, we finally go back to the ferromagnetic resonance case and we use SWIIM to calculate the resonance frequencies of the two lowest FMR modes as a function of magnetic field using the now completed set of magnetic parameters (Table~\ref{tab:table2}). Then we fit the frequencies determined numerically to Eqs.~(\ref{eq:p1}) and extract the corresponding stiffness fields. As expected, accounting for the difference in surface anisotropy evidenced through propagating spin wave spectroscopy allows one improving slightly the agreement between experimentally and numerically determined $H_{Y0}$, $H_{Z0}$, $H_{Y1}$, and $H_{Z1}$ stiffness fields (Fig.~\ref{fig:HiFields}). Note that despite the introduction of $\Delta K_{\text{S}}$ and the exact treatment of hybridization effects by SWIIM, numerical data remain however rather close to predictions of our analytical model. This proves the suitability of our choice of a limited four-vector mode basis (Sec.~\ref{sec:level2_Model}).

\section{\label{sec:level4_Discussion}Discussion}
Starting from a simplified analytical model, we have described above a method for extracting the magnetic parameters of ferromagnetic films with moderate thickness from broadband ferromagnetic resonance and propagating spin wave spectroscopy measurements. The positions of the ferromagnetic resonance peaks measured over large field and frequency ranges are first fitted to Kittel formulas (Fig.~\ref{fig:FMRfvsH}). Then, the extracted stiffness fields are confronted to explicit expressions (Fig.~\ref{tab:table1}) allowing one to extract successively five magnetic parameters. The deviations between the model and the experiments do not exceed 3$\%$ (Fig.~\ref{fig:HiFields}), which we find very satisfactory given the wide range of field, frequency and film thickness investigated, and the limited number of parameters involved. Moreover, the extension of the model to propagating spin waves accounts for the measured frequency non-reciprocity, a quantity from which we extract a sixth magnetic parameter. The value of the latter is eventually refined by confronting frequency non reciprocity data to full micromagnetic calculations.

The above ferromagnetic resonance study suggests that the entire thickness series can be described quite accurately with a thickness independent set of magnetic parameters. The exchange stiffness constant we determined lies within the range of values reported in literature for Fe at room temperature $A\!=\!19\text{-}23~\text{pJ/m}$~\cite{Devolder.2013,Niitsu.2020,Kuzmin.2020} and agrees particularly well with the recent determinations by Niitsu~\cite{Niitsu.2020} and Kuz'min~\textit{et al.} \cite{Kuzmin.2020}. The measured cubic anisotropy constant, on the other hand, is slightly larger than the values measured on bulk iron and thin iron films ($K_1\!=\!44\text{-}49~\text{kJ/m}^{3}$) \cite{Buschow.2001,Graham.1958,Westerstrand.1975,Barsukov.2011}, but it agrees well with results from first principles calculations ($K_1\!=\!52~\text{kJ/m}^{3}$)~\cite{Barsukov.2011}.

The third volume parameter, namely the uniaxial anisotropy $K_{\text{U}}$, accounts for the difference between the saturation magnetization $M_{\text{s}}$ determined from SQUID magnetometry and the so-called effective magnetization $M_{\text{eff}}\!=\!(M_{\text{s}}\!-\!H_{\text{U}})$ measured from FMR. Now, we argue that this anisotropy originates from a distortion of the iron lattice. Indeed, the -4$\%$ mismatch between Fe and the MgO substrate is known to relax only partly through a dense array of dislocations formed in the first Fe atomic layers, thus leaving a small residual strain in nanometer thick films~\cite{Du2021}. Accordingly, a tetragonal distortion is measured in the samples under study consisting of a 0.7$\%$ mean in-plane expansion and a 0.5$\%$ out-of-plane compression~\cite{Magnifouet.2020}. Such vertical lattice compression is expected to enhance the spin-orbit mediated interactions between electronic states which favor an in-plane orientation of the magnetization~\cite{Wu.1998}. To relate phenomenologically this extra magnetic anisotropy  to the measured lattice distortion one may use the magnetoelastic coupling constants of bulk iron~\cite{Hearmon.1946}. The obtained uniaxial anisotropy constant $K_{\text{U}}\!=\!-41$~kJ/m$^{3}$ (see Sander~\cite{Sander.2004} for calculation details) is in very good agreement with our experimental observations regarding both its sign and its magnitude.

In terms of total perpendicular surface anisotropy, our results (see Table~\ref{tab:table2}) agree well with what is expected for an Fe ultra thin film sandwiched between two MgO layers~\cite{Shimabukuro.2010,Nakamura.2010,Yang.2011, Hallal.2013, Odkhuu.2016}. From the joint results of broadband FMR and spin wave propagation, we can extract the values for the two individual PSA constants: $K^{\text{bot}}_{\text{S}}\!=\!(K_{\text{S}}+\Delta K_{\text{S}})/2\!=\!1.55$~mJ/m$^{2}$ for the bottom interface (MgO buffer/Fe) and $K^{\text{top}}_{\text{S}}\!=\!(K_{\text{S}}-\Delta K_{\text{S}})/2\!=\!0.75$~mJ/m$^{2}$ for the top one (Fe/MgO capping). These two values are within the range for the PSA obtained by ab-initio calculations~\cite{Shimabukuro.2010,Nakamura.2010,Yang.2011, Hallal.2013, Odkhuu.2016} and measurements on ultrathin films with a single MgO/Fe interface~\cite{Lambert.2013,Koo.2013,Okabayashi.2014,KozioRachwa.2013,KozioRachwa.2013b}. However, in these previous experimental works, the extracted values always included the contributions from two interfaces, and some hypothesis based on reference interfaces (e.g. V/Fe) needed to be included to extract individual values. In the present study we provide individual $K_{\text{S}}$ values for both interfaces which allows us to compare them directly with results from ab-initio simulations and evidence that the ultra-thin interface physics can be extrapolated to thicker films~\cite{Shimabukuro.2010,Nakamura.2010,Yang.2011, Hallal.2013, Odkhuu.2016, Lambert.2013,Koo.2013,Okabayashi.2014,KozioRachwa.2013,KozioRachwa.2013b}.

It has been shown theoretically that over/under oxidation at the Fe/MgO interface reduce drastically its surface anisotropy~\cite{Yang.2011,Hallal.2013}. Moreover, using Mossbaüer spectroscopy, it has been shown that the Fe/MgO and MgO/Fe interfaces of a film can exhibit different amount of interfacial Fe oxidation~\cite{Mynczak.2013}. Therefore, we attribute the difference in PSA at the two interfaces to a difference in their oxidation states. We assume that the distinct temperature treatments to which the bottom and top interfaces are subjected during growth is the reason for that: the 480°C annealing, performed just after iron deposition, promotes a better epitaxy and higher value of surface anisotropy for the bottom MgO/Fe interface~\cite{Okabayashi.2014}. On the other hand, the top interface is not annealed, which likely leads to an over oxidation of the interfacial Fe atoms and a lower value of the PSA~\cite{Yang.2011,Hallal.2013}. This behaviour is corroborated by the larger value of $\Delta K_{\text{S}}$ observed in the 20~nm* sample without Ti protection (see blue dots in Fig.~\ref{fig:NRvsk}). For this sample, indeed, we argue that a further oxidation of the top interface may take place after the unprotected 8~nm thick MgO capping layer is exposed to water~\cite{Holt.1997,Youssef.2018} both during the fabrication of this specific device and later under ambient conditions.

\section{Conclusion}
The magnetization dynamics of a thickness series of MgO/Fe($t$)/MgO epitaxial films ($t\!\text{=}\!10\text{-}30$~nm) was characterized using a combination of ferromagnetic resonance and non-reciprocal spin wave propagation measurements. Our rather versatile Kittel model accounts consistently for the frequencies of the uniform mode of magnetization precession and also for the inhomogeneous first standing spin-wave mode. Noticeably, the ability to probe both of these modes over a wide range of film thicknesses allowed us to determine the exchange stiffness constant and the perpendicular surface anisotropy, two quantities which are inaccessible through the sole study of homogeneous dynamics.

With our detailed ferromagnetic resonance characterization, we evidenced that the entire thickness series can be described with a single set of magnetic parameters. The magnetic volume parameters, cubic anisotropy and exchange stiffness, agree very well with what is expected for bulk iron. Also, an additional uniaxial perpendicular anisotropy was identified and attributed to a slight tetragonal distortion of the Fe lattice. Finally, it was possible to separate contributions of individual film interfaces to the perpendicular surface anisotropy with the help of complementary propagating spin wave spectroscopy measurements. The sizeable asymmetry between the top and bottom interfaces was attributed to the different oxidation states of each interface. Our characterization suggest that 10-30 nm thick single crystalline Fe films have a well defined quasi-bulk magnetic interior, while the interfaces with MgO retain the large perpendicular surface anisotropy found in ultra-thin film. 

Our work provides new light into the technologically-relevant ferromagnet/MgO interfaces and their effect on spin waves, while it also validates a new method for characterizing magnetic interfaces. Note that our methodology could be extended to alloys or multilayer systems, for which it could provide key information about possible inhomogeneities / asymmetries of the magnetic properties across the film thickness. 

\begin{acknowledgments}

We thank Arnaud Boulard, Beno\^{\i}t Leconte, Daniel Spor, J\'er\'emy Thoraval and Fares Abiza for assembling and testing the broadband FMR setup; J\'er\^{o}me Robert for SQUID magnetometry measurements; Romain Bernard, Sabine Siegwald and Hicham Majjad for technical support during nanofabrication work in the STnano platform; and Mat\'{\i}as Grassi for useful discussion. We acknowledge financial support by the Interdisciplinary Thematic Institute QMat, as part of the ITI 2021-2028 program of the University of Strasbourg, CNRS and Inserm, IdEx Unistra (ANR 10 IDEX 0002), SFRI STRAT’US project (ANR 20 SFRI 0012) and ANR-17-EURE-0024 under the framework of the French Investments for the Future Program. We also acknowledge financial support from Region Grand Est through its FRCR call (NanoTeraHertz and RaNGE projects) and from Agence Nationale de la Recherche (France) under Contract No. ANR-20-CE24-0012 (MARIN).

\end{acknowledgments}


\appendix

\section{\label{sec:App1}Dynamic matrices}

To find the resonance frequencies of the magnetization modes one has to solve the Landau-Lifshitz (LL) equation. For this, we substitute Eq.~(\ref{eq:StaticField}) for the static field, Eq.~(\ref{eq:PropDynH}) for the dynamic field, and Eq.~(\ref{eq:Dynamicmag}) for the dynamic magnetization into the linearized LL Eq.~(\ref{eq:LL}) and project the latter onto the space of cosine thickness modes, as explained in Section~\ref{sec:level2_FMR}~\cite{Kalinikos.1986}. To be able to derive useful analytical solutions, it is convenient to restrict this projection to the first two modes, yielding a total of four basis vectors $\left[\Phi_0\mathbf{\hat{x}},\;\Phi_1(\xi)\mathbf{\hat{x}},\; \Phi_0\mathbf{\hat{y}},\; \Phi_1(\xi)\mathbf{\hat{x}}\right]$ (two modes per spacial coordinate of the dynamic magnetization). After projection, the linearized LL equation [Eq.~(\ref{eq:LL})] takes the form of an eigenvalue equation: $i\omega \Bar{\textbf{m}}^*=C\Bar{\textbf{m}}^*$, where $C$ is the so called dynamic matrix, which in the present case, is 4$\times$4~\cite{Gladii.2016p}. The eigenvalues of this matrix are the resonance frequencies and the eigenvectors describe the corresponding dynamic magnetization mode amplitudes. In Eq.~\ref{eq:Cmatrix}, we provide expressions for the dynamic matrices in cases where the external magnetic field is applied in-plane ($C_\parallel$) and out-of-plane ($C_\perp$).
\begin{widetext}
\begin{subequations}
\label{eq:Cmatrix}
\begin{equation}
{
C_\parallel\text{=}\gamma \mu_{\text{o}}
\begin{pmatrix}
0  &  H_\parallel(k)\text{+}M_{\text{s}} P_{00}(k) & -i  M_{\text{s}} Q(k) & 0 \\
 \text{-}[H_\parallel(k)\text{-}H_{\text{U}}\text{-}H_{\text{S}}\text{+}M_{\text{s}}[1\text{-}P_{00}(k)]] & 0  & \text{-}\frac{2\sqrt{2}\Delta K_{\text{S}}}{\mu_{\text{o}} t M_{\text{s}} } & i  M_{\text{s}} Q(k) \\
 i  M_{\text{s}} Q(k) & 0 & 0  & H_\parallel(k)\text{+}H_{\text{E}}\text{+}M_{\text{s}} P_{11}(k) \\
 \text{-}\frac{2\sqrt{2}\Delta K_{\text{S}}}{\mu_{\text{o}} t M_{\text{s}} } & \text{-}i  M_{\text{s}} Q(k) & \text{-}[H_\parallel(k)\text{+}H_{\text{E}}\text{-}H_{\text{U}}\text{-}2H_{\text{S}}\text{+}M_{\text{s}}[1\text{-}P_{11}(k)]] & 0 \\
\end{pmatrix}
\label{subeq:3}
}
\end{equation}
\begin{equation}
{
C_\perp\text{=}\gamma \mu_{\text{o}}
\begin{pmatrix}
0  &  H_\perp(k)\text{+}M_{\text{s}} P_{00}(k) & 0 & \text{-}\frac{2\sqrt{2}\Delta K_{\text{S}}}{\mu_{\text{o}} t M_{\text{s}} } \\
 \text{-}H_\perp(k) & 0  & \frac{2\sqrt{2}\Delta K_{\text{S}}}{\mu_{\text{o}} t M_{\text{s}} } & 0 \\
 0 & \text{-}\frac{2\sqrt{2}\Delta K_{\text{S}}}{\mu_{\text{o}} t M_{\text{s}} } & 0  & H_\perp(k)\text{+}H_{\text{E}}\text{+}H_{\text{S}}\text{+}M_{\text{s}} P_{11}(k) \\
 \frac{2\sqrt{2}\Delta K_{\text{S}}}{\mu_{\text{o}} t M_{\text{s}} } & 0 & -[H_\perp(k)\text{+}H_{\text{E}}\text{+}H_{\text{S}}] & 0   \\
\end{pmatrix}.
\label{subeq:4}
}
\end{equation}
\end{subequations}
\end{widetext}

\noindent The fields $H_\parallel(k)\text\!=\!H\!+\!H_{\text{K}}\!+\!M_{\text{s}}\Lambda^2 k^2$ and $H_\perp(k)\!=\!H+H_{\text{K}}\!+\!H_{\text{U}}\!+\!H_{\text{S}}+M_{\text{s}} (\Lambda^2 k^2$-1) are intermediate parameters introduced to simplify those expressions. $P_{00}(k)\!=\!1\!-\!\frac{1-e^{-|k| t}}{|k| t }$ and $P_{11}(k)\!=\!\frac{(k t)^2}{\pi^2+(kt)^2}\left(1\!-\!\frac{2(kt)^2}{\pi^2+(kt)^2}\frac{1+e^{-|k|t}}{|k|t}\right)$ are self demagnetizing factors for the cosine thickness modes $n\!=\!0$ and $n\!=\!1$, and $Q\!=\!\frac{\sqrt{2}kt}{\pi^2+(kt)^2}(1+e^{-|k|t})$ is a mutual demagnetizing factor responsible for hybridization between those modes; all three factors being part of the magnetostatic Green's function that describes the dipolar interaction~\cite{GUSLIENKO.2011}.

\section{\label{sec:App3}Effect of a non-zero $\Delta K_{\text{S}}$ on the stiffness fields in the case of ferromagnetic resonance}

In Section~\ref{sec:level2_FMR} we have derived Eq. (\ref{eq:p1}) for the resonance frequencies of the modes $n\!=\!0,1$ by considering the effect of $\Delta K_{\text{S}}$ only up to first order. In the infinite wavelength limit ($k\!=\!0$), those frequencies become fully independent of $\Delta K_{\text{S}}$. Now, we proceed to consider the approximation up to second order in this parameter and study its effect on Eq.~(\ref{eq:p1}).

When keeping terms proportional to $\Delta K_{\text{S}}^2$ in the model, the stiffness fields need to be modified as follows: $H_{\text{X}n} \to H_{\text{X}n}$, $H_{\text{Y}n} \to H_{\text{Y}n}+g_{\text{Y}n}$, $H_{\text{Z}0} \to H_{\text{Z}0} + g_{\text{Z}}$, and  $H_{\text{Z}1} \to H_{\text{Z}1}-g_{\text{Z}}$, where the field corrections $g_{\text{Y}n}$ ($n\!=\!0,1$) and $g_{\text{Z}}$ are given by 
\begin{subequations}
\label{eq:Corrections}
\begin{equation}
{
g_{Y0}(H,t)=\frac{-8\frac{ H_\text{E}+H_\text{K}+H }{({\mu_0 M_\text{s} t)}^2}\Delta K_\text{S}^2}{\scriptstyle{H_\text{S}(H_\text{K}+H)+H_\text{E}[H_\text{U}-M_\text{s}-H_\text{E}+2(H_\text{S}-H_\text{K}-H)]}}
\label{subeq:7}
}
\end{equation}
\begin{equation}
{
g_{Y1}(H,t)=\frac{8\frac{H_\text{K}+H }{({\mu_0 M_\text{s} t)}^2}\Delta K_\text{S}^2 }{\scriptstyle{H_\text{S}(H_\text{K}+H)+H_\text{E}[H_\text{U}-M_\text{s}-H_\text{E}+2(H_\text{S}-H_\text{K}-H)]}}
\label{subeq:8}
}
\end{equation}
\begin{equation}
{
g_{Z}(t)=\frac{8\Delta K_\text{S}^2}{({\mu_0 M_\text{s} t)}^2(H_\text{E}+H_\text{S})}\text{.}
\label{subeq:9}
}
\end{equation}
\end{subequations}

\begin{figure}[ht]
\includegraphics[width=86mm]{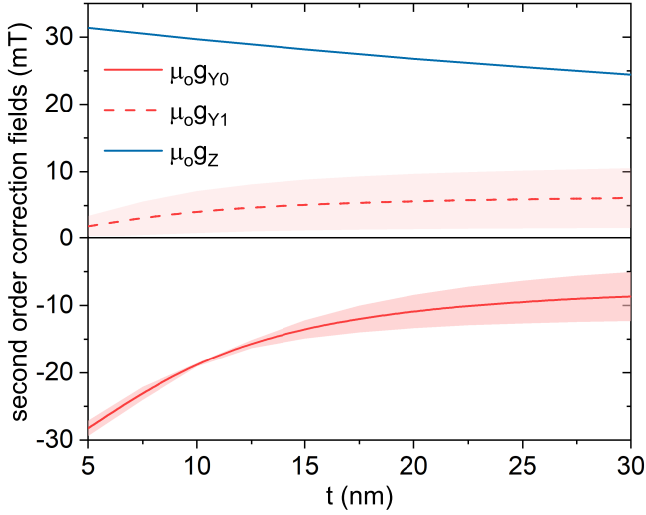}%
\caption{\label{fig:2nd} Second order correction functions for the stiffness fields (Eq.~\ref{eq:Corrections}) as calculated with $\Delta K_{\text{S}}=1.65$~mJ/m$^2$ to obtain a sizable effect. Functions $g_{Yn}$ and $g_{\text{Z}}$ are the stiffness field corrections for in-plane applied field, while $g_{\text{Z}}$ applies to the out-of-plane case. The shaded zones around the lines account for the variations of $g_{\alpha n}$ with applied magnetic field in the range 0-1.3~T.}
\end{figure}

\noindent $g_{\text{Z}}$ is completely independent of the external field $H$ and $g_{\text{Y}n}$ depends only very weakly on it in the range of interest (Fig.~\ref{fig:2nd}) so that it can also be considered as constant. Therefore, under this second order approximation, the resonance frequencies $f_{\parallel n}$ and $f_{\perp n}$ retain approximately the Kittel-like form of Eq.~\ref{eq:p1}, with the external field dependence contained entirely in the explicit $H$ term. This justifies our fitting of the ferromagnetic resonance frequencies in Fig.~\ref{fig:FMRfvsH} to Eq.~\ref{eq:p1} even in the presence of a sizable $\Delta K_{\text{S}}$. We note in passing that, given their smallness (Fig.~\ref{fig:2nd}), the correction fields will have a negligible effect on the determination of the main parameters $K_{\text{S}}$ and $K_{\text{U}}$.

\begin{figure}[ht]
\includegraphics[width=86mm]{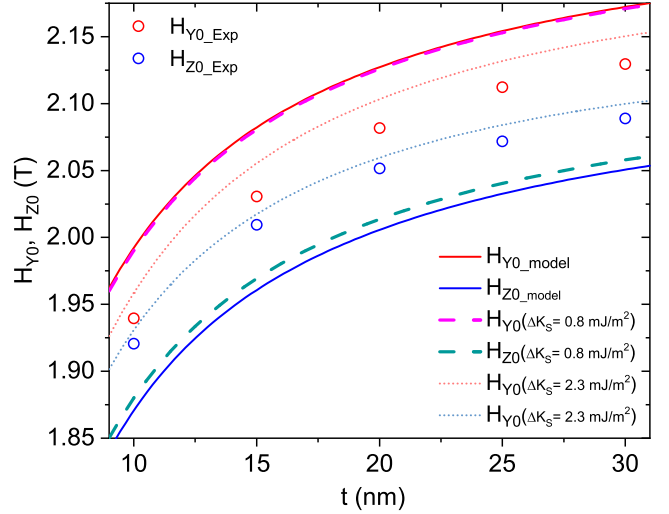}%
\caption{\label{fig:Hy0Hz02nd}Stiffness fields $H_{\text{Y}0}$ (red) and $H_{\text{Z}0}$ (blue) as functions of the Fe film thickness $t$. The open circles are the experimentally determined values and the lines are predictions of our analytical approach. Continuous, dashed, and dotted lines correspond, respectively, to first order approximation, second order approximation with $\Delta K_{\text{S}}\!=\!0.8$~mJ/m$^{2}$, and second order approximation with $\Delta K_{\text{S}}\!=\!2.3$~mJ/m$^{2}$.}
\end{figure}

As mentioned in the body of the paper (Sec.~\ref{sec:results}), moving to second order approximation allows one to improve qualitatively the agreement between theoretical and experimental values of the stiffness fields (Fig.~\ref{fig:Hy0Hz02nd}). However, when assuming the value $\Delta K_{\text{S}}\!=\!0.8$~mJ/m$^{2}$ determined from spin-wave spectroscopy (Sec.~\ref{sec:level3_PSWS}) the improvement remains marginal, especially regarding $H_{Y0}$, and only a much bigger value of $\Delta K_{\text{S}}$ allows one reaching a reasonably good matching (note the different vertical scales in Figs.~\ref{fig:HiFields} and \ref{fig:Hy0Hz02nd}). Here again, we evidence the tendency of our analytical approach to underestimate the effect of a difference in surface anisotropies. We attribute this quantitative discrepancy partially to the hypothesis made in writing Eq.~\ref{eq:Dynamicmag}, which is to neglect higher order terms in the Fourier series. We overcome this limitation with our numerical analysis (SWIMM code), which allows us to obtain a better overall agreement with broadband FMR and PSWS (see dotted and dashed lines in Figs.~\ref{fig:HiFields} and \ref{fig:Hy0Hz02nd}).

\FloatBarrier
\providecommand{\noopsort}[1]{}\providecommand{\singleletter}[1]{#1}%
%
\end{document}